\documentclass[preprint,twocolumn]{aastex631}

\usepackage[utf8]{inputenc}
\usepackage{graphicx}
\usepackage{epsfig}
\usepackage{natbib}
\usepackage{footnote}
\usepackage{url}
\usepackage{appendix}

\graphicspath{{./}{figures/}}

\submitjournal{(and accepted) RNAAS}

\begin{document}

\title{Do the Outburst Properties of M31N 2008-12a Depend on the Time Since the Previous Eruption?}

\author[0000-0002-6023-7291]{William A. Burris}
\affiliation{Department of Astronomy, San Diego State University, 5500 Campanile Drive, San Diego, CA 92182, USA}

\author[0000-0002-1276-1486]{Allen W. Shafter}
\affiliation{Department of Astronomy, San Diego State University, 5500 Campanile Drive, San Diego, CA 92182, USA}

\author[0000-0002-0835-225X]{Kamil Hornoch}
\affiliation{Astronomical Institute of the Czech Academy of Sciences, Fri\v{c}ova 298, CZ-251 65 Ond\v{r}ejov, Czech Republic}

\begin{abstract}
 Photometric observations spanning the UV to the near IR during the nine most recent eruptions (2014--2022) of the extragalactic nova M31N 2008-12a are presented and analyzed in order
 to explore whether the lightcurve properties for a given eruption, specifically the peak magnitudes and fade rates, are correlated with the time interval since the previous eruption. No significant correlation between the pre-eruption interval and the rate of decline was found, however it appears that the brightness at the peak of an outburst may be positively correlated with the time interval since the previous eruption.
 
\end{abstract}

\keywords{Andromeda Galaxy (39) -- Novae (1127) -- Recurrent Novae (1366)}

\section{Introduction}
M31N 2008-12a was discovered by
Koichi Nishiyama and Fujio Kabashima, on 2008 Dec. 26.48~UT in the outskirts of M31\footnote{\url{http://www.cbat.eps.harvard.edu/iau/CBAT\_M31.html\#2008-12a}}.
After the object was seen again in the fall of 2011 and 2012, \citet{2012ATel.4503....1S} speculated that the object was either a recurrent nova or a slow nova undergoing multiple rebrightenings. When the object erupted again in the fall of 2013, the recurrent nova nature of the object became clear \cite[][]{2014A&A...563L...9D,2014ApJ...786...61T}.
Based on timings of the 15 eruptions observed every year since 2008, the mean recurrence time is $364.18\pm2.18$~d, or $0.997\pm0.006$~yr, which is the shortest of any known nova.

The $\sim1$~yr recurrence time offers an exceptional opportunity to study how the properties of a nova progenitor affects the outburst behavior. In the case of 2008-12a, the short recurrence time requires that the white dwarf in the system must be near the Chandrashekhar mass and accreting at a rate a few times $10^{-7}~M_{\odot}$~yr$^{-1}$ \citep[e.g.,][]{2014ApJ...793..136K}.

Starting in 2014 a worldwide campaign to monitor 2008-12a has produced a wealth of photometric data \cite[e.g., see][and references therein]{2020AdSpR..66.1147D}.
Here, we analyze lightcurve data from 2014 to the most recent eruption in 2022.
Our primary goal is to explore whether properties of the UV and optical lightcurves are affected by the time since the previous eruption ($t_\mathrm{pre}$).

\begin{figure*}
    \centering
    \includegraphics[scale=0.65]{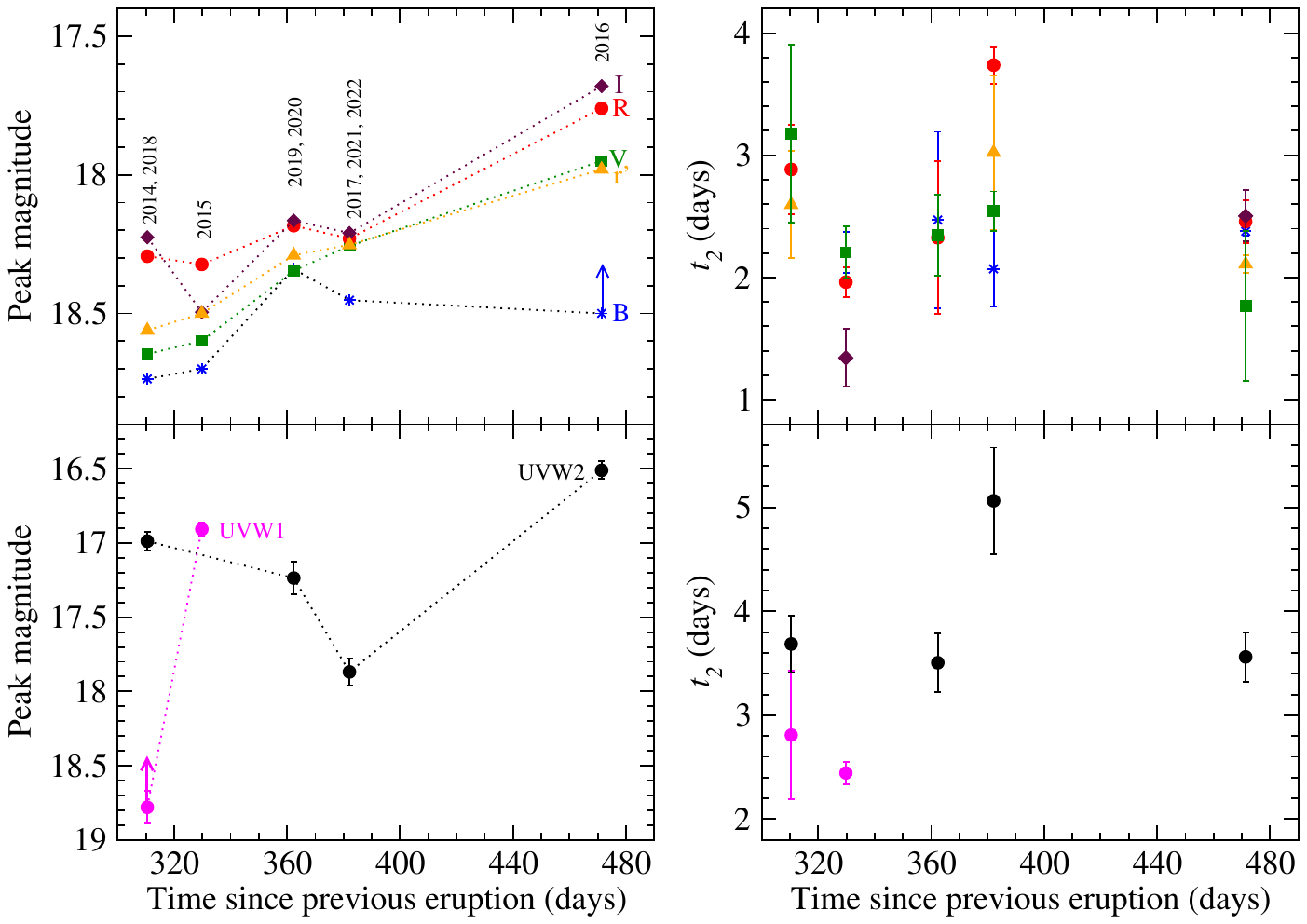}
    \caption{{\it Top Left}: The peak optical magnitudes reached by 2008-12a as a function of $t_\mathrm{pre}$. For values of $t_\mathrm{pre}$ representing multiple eruptions, the average peak magnitude is shown. A likely trend of peak brightness with $t_\mathrm{pre}$ is evident. The $B$ peak magnitude for the 2016 eruption, which was obtained $\sim$5~hr after the other filters, may have faded somewhat during that interval.
    {\it Bottom Left}: Peak magnitude as a function of $t_\mathrm{pre}$ for the {\it Swift\/} UVOT bandpasses. No clear trend is observed, likely because the data were typically obtained after optical identification. {\it Top (Bottom) Right}: The $t_2$ time as a function of $t_\mathrm{pre}$ for the optical (UVOT) bandpasses. In both cases no correlation between $t_2$ time and $t_\mathrm{pre}$ is observed. (See Appendix for DBF)}
    \label{fig:1}
\end{figure*}

\section{Photometric Data}

Optical (Johnson $B,V,R,I$ and Sloan $r$) and UVOT (UVW1 and UVW2) photometry for eruptions from 2014 to 2022 have been analyzed to determine lightcurve parameters. The optical data were gleaned from the literature, while the UVOT photometry was obtained from the {\it Swift\/} archive. The UVOT data were reanalyzed using the {\it HEASoft} tool {\it uvotmaghist} to achieve the largest temporal coverage available, while using smaller apertures ($3''$ instead of the $5''$ used in the default {\it Swift\/} calibration) and multiple background regions to maximize the signal-to-noise.
Unfortunately, the UVOT data were not always obtained using the same filters making direct comparisons between the various eruptions difficult. Both the UVW1 ($\lambda_\mathrm{eff} \sim 2600$ \AA) and UVW2 ($\lambda_\mathrm{eff} \sim 1928$ \AA) filters were used in 2014, while in subsequent years only one filter was employed: UVW1 in 2015, and UVW2 from 2016 onwards.

\section{LightCurve Parameters}

Linear least-square fits to $\sim$2 mag below peak were performed on the rise (when observed) and fall from maximum light. When the rise was observed, the intersection of the two linear fits was taken to be the time of maximum light, and the extrapolated magnitude at that time was taken to be the peak magnitude. When the rise was not captured, we took the brightest observed point to represent
maximum light\footnote{Formally, this represents a lower limit to the actual peak magnitude, but due to the sustained monitoring around the world, observations were typically made within hours of discovery, and thus likely close to peak brightness.}. In all cases, the rate of decline from maximum light was used to determine the corresponding $t_2$ times (the time to decline by 2 mag from maximum light) for each eruption. There were some years where only a few data points in a particular filter were measured, not enough to measure a meaningful $t_2$ time. In these cases, we also adopted the brightest measured magnitude as maximum light.

The peak magnitudes and $t_2$ times are shown in Figure~\ref{fig:1} where we have plotted the parameters
as a function of the time interval since previous eruption, $t_\mathrm{pre}$. Given that $t_\mathrm{pre}$ was (remarkably!) nearly identical for several epochs, we averaged the maximum magnitudes and $t_2$ times for those pre-eruption intervals. This process resulted in five mean intervals of $\langle t_\mathrm{pre}\rangle$ = 310.49~d (2014/2018), 329.83~d (2015), 362.34~d (2019/2020), 382.09~d (2017/2021/2022), and 471.34~d (2016).

No correlation is apparent between the $t_2$ and $t_\mathrm{pre}$ times, however we do find tentative evidence for a correlation between the peak brightness and $t_\mathrm{pre}$. The most compelling evidence is provided by the 2016 eruption, which had the longest pre-eruption interval of $t_\mathrm{pre}=471.34$ d. In 4 out of the 5 optical filters, and in the UVW2 filter, the peak brightness of the 2016 eruption is largest seen, even though the peak was not captured in any filter that year. On the other hand, the 2014 and 2018 eruptions, which had the shortest pre-eruption mean interval ($\langle t_\mathrm{pre}\rangle = 310.49$ d), were the faintest of the nine most recent eruptions (in all but the $I$ and $R$ filters), although the 2018 lightcurve morphology was somewhat complicated
exhibiting a double-peaked maximum: a dip after the original peak followed by a short resurgence. It is unclear whether the double-peak structure of the outburst is related to the short pre-eruption interval, but it will be interesting to see if 2008-12a exhibits similar behavior during future eruptions
with short $t_\mathrm{pre}$. There is no obvious double-peaked structure in the 2014 eruption, but the data are limited.

\section{Conclusions}
\label{sec:5}
Our analysis has failed to reveal any obvious relationship between the $t_2$ time and $t_\mathrm{pre}$. However, we did find tentative evidence that the peak optical brightness reached by M31N 2008-12a increases with the time since the previous eruption. The trend is not obvious in the UV likely because the
{\it Swift\/} observations were always triggered after the eruption was discovered in the optical giving the nova more time to fade before the UV photometry could be performed.

The accreted mass required to trigger an eruption depends on $M_\mathrm{WD}$ and the mean accretion rate onto the white dwarf's surface \citep[e.g.,][]{2005ApJ...628..395T}. A longer $t_\mathrm{pre}$ implies a lower $\langle dM/dt \rangle$ between eruptions. The observed trend of peak brightness with $t_\mathrm{pre}$ may be the result of increased degeneracy in the accreted layer that would be expected for this slightly lower accretion rate.
Continued monitoring of 2008-12a during future eruptions will be required in order to confirm this finding, and if warranted, to explore its origin.

\begin{acknowledgements}
This work has been supported by NASA grant 80NSSC20K0547 (AWS) and by the project RVO:67985815 (KH). We thank K. Page for her advice on the {\it Swift\/} reductions.
\end{acknowledgements}

\facilities{{\it Swift\/}: https://swift.gsfc.nasa.gov/archive/}

\bibliography{M31N2008-12a.bib}

\appendix

\startlongtable
\begin{deluxetable}{ccccr}
\tablenum{A1}
\tablecolumns{5}
\tablecaption{M31N 2009-12a Lightcurve Parameters (Figure 1)}
\tablehead{\colhead{$t_\mathrm{pre}$ (d)} & \colhead{Mag} & \colhead{$t_2$ (d)} & \colhead{Filter} & \colhead{Source\tablenotemark{a}}}
\startdata
310.35 &$ 17.20\pm 0.07$&$  3.96\pm 0.51$&$    UVW2$&  This work \cr
310.35 &$ 18.83\pm 0.01$&$  \dots       $&$    B   $&   1,2,3,4 \cr
310.35 &$ 18.79\pm 0.01$&$  3.18\pm 0.73$&$    V   $&   1,2,3,4 \cr
310.35 &$ 18.35\pm 0.08$&$  2.49\pm 0.34$&$    R   $&   1,2,3,4 \cr
310.35 &$ 18.66\pm 0.02$&$  2.60\pm 0.44$&$    r'  $&   4,5 \cr
310.35 &$ 18.21\pm 0.05$&$  \dots       $&$    I   $&   1,2,3,4 \cr
310.64 &$ 18.78\pm 0.11$&$  2.81\pm 0.62$&$    UVW1$&  This work \cr
310.64 &$ 16.81\pm 0.06$&$  3.45\pm 0.19$&$    UVW2$&  This work \cr
310.64 &$ 18.65\pm 0.15$&$  \dots       $&$    B   $&  6,7 \cr
310.64 &$ 18.52\pm 0.10$&$  \dots       $&$    V   $&  6,7 \cr
310.64 &$ 18.24\pm 0.10$&$  3.42\pm 0.40$&$    R   $&  6,7 \cr
310.64 &$ 18.47\pm 0.01$&$  \dots       $&$    r'  $&  8 \cr
310.64 &$ 18.24\pm 0.11$&$  \dots       $&$    I   $&  6,7 \cr
329.83 &$ 16.91\pm 0.04$&$  2.44\pm 0.11$&$    UVW1$&  This work \cr
329.83 &$ 18.70\pm 0.01$&$  2.20\pm 0.17$&$    B   $&  9,10,11,12,13,14,15 \cr
329.83 &$ 18.60\pm 0.03$&$  2.21\pm 0.22$&$    V   $&  9,10,11,12,13,14,15,16,17,18        \cr
329.83 &$ 18.32\pm 0.02$&$  1.96\pm 0.12$&$    R   $&  10,11,17 \cr
329.83 &$ 18.50\pm 0.02$&$  \dots       $&$    r'  $&  9,12,13 \cr
329.83 &$ 18.50\pm 0.08$&$  1.34\pm 0.24$&$    I   $&  10,11,17 \cr
359.37 &$ 17.46\pm 0.14$&$  3.71\pm 0.71$&$    UVW2$&  This work \cr
359.37 &$ 18.38\pm 0.08$&$  \dots       $&$    R   $&  This work \cr
359.37 &$ 18.29\pm 0.07$&$  \dots       $&$    r'  $&  19,20  \cr
359.37 &$ 18.15\pm 0.10$&$  \dots       $&$    I   $&  This work \cr
365.31 &$ 17.05\pm 0.08$&$  3.32\pm 0.18$&$    UVW2$&  This work \cr
365.31 &$ 18.34\pm 0.16$&$  2.47\pm 0.72$&$    B   $&  21,22,23 \cr
365.31 &$ 18.35\pm 0.05$&$  2.35\pm 0.33$&$    V   $&  21,22,23 \cr
365.31 &$ 18.02\pm 0.05$&$  2.33\pm 0.63$&$    R   $&  21,22,23 \cr
365.31 &$ 18.18\pm 0.03$&$  \dots       $&$    I   $&  21,23  \cr
379.18 &$ 17.69\pm 0.09$&$  5.42\pm 0.60$&$    UVW2$&  This work \cr
379.18 &$ 18.44\pm 0.05$&$  \dots       $&$    B   $&  24  \cr
379.18 &$ 18.39\pm 0.06$&$  3.22\pm 0.34$&$    V   $&  24,25,26 \cr
379.18 &$ 18.25\pm 0.04$&$  2.40\pm 0.07$&$    R   $&  This work \cr
379.18 &$ 18.28\pm 0.06$&$  \dots       $&$    r'  $&  24,25,26,27 \cr
383.16 &$ 18.02\pm 0.09$&$  4.11\pm 0.46$&$    UVW2$&  This work \cr
383.16 &$ 18.52\pm 0.06$&$  3.19\pm 0.17$&$    B   $&  28 \cr
383.16 &$ 18.35\pm 0.27$&$  3.12\pm 0.20$&$    V   $&  28 \cr
383.16 &$ 18.26\pm 0.01$&$  6.21\pm 0.33$&$    R   $&  28 \cr
383.16 &$ 18.23\pm 0.10$&$  3.02\pm 0.63$&$    r'  $&  29,30,31,32,33,34     \cr
383.98 &$ 17.92\pm 0.09$&$  6.08\pm 0.50$&$    UVW2$&  This work \cr
383.98 &$ 18.40\pm 0.25$&$  2.31\pm 0.26$&$    B   $&  35,36,37   \cr
383.98 &$ 18.06\pm 0.09$&$  1.83\pm 0.09$&$    V   $&  35,36,37,38   \cr
383.98 &$ 18.18\pm 0.09$&$  4.44\pm 0.38$&$    R   $&  35,36,37,38   \cr
383.98 &$ 18.21\pm 0.10$&$  \dots       $&$    I   $&  36,37,38      \cr
471.37 &$ 16.51\pm 0.06$&$  3.56\pm 0.24$&$    UVW2$&  This work \cr
471.37 &$ 18.50\pm 0.10$&$  2.38\pm 0.04$&$    B   $&  39            \cr
471.37 &$ 17.95\pm 0.09$&$  1.77\pm 0.61$&$    V   $&  39,40      \cr
471.37 &$ 17.76\pm 0.05$&$  2.46\pm 0.18$&$    R   $&  39,40,41  \cr
471.37 &$ 17.98\pm 0.04$&$  2.11\pm 0.07$&$    r'  $&  40,41,42  \cr
471.37 &$ 17.68\pm 0.08$&$  2.50\pm 0.21$&$    I   $&  39,40    \cr
\enddata
\tablenotetext{a}{
1: ATel \#12181,
2: ATel \#12190,
3: ATel \#12200,
4: ATel \#12205,
5: ATel \#12203,
6: ATel \#6543,
7: ATel \#6546,
8: ATel \#6535,
9: ATel \#7965,
10: ATel \#7967,
11: ATel \#7969,
12: ATel \#7980,
13: ATel \#7984,
14: ATel \#8029,
15: ATel \#8038,
16: ATel \#7974,
17: ATel \#7976,
18: ATel \#7979,
19: ATel \#14130,
20: ATel \#14131,
21: ATel \#13279,
22: ATel \#13281,
23: ATel \#13302,
24: ATel \#15037,
25: ATel \#15039,
26: ATel \#15068,
27: ATel \#15038,
28: ATel \#15797,
29: ATel \#15786,
30: ATel \#15788,
31: ATel \#15789,
32: ATel \#15795,
33: ATel \#15802,
34: ATel \#15902,
35: ATel \#11118,
36: ATel \#11125,
37: ATel \#11126,
38: ATel \#11124,
39: ATel \#9864,
40: ATel \#9861,
41: ATel \#9857,
42: ATel \#9881.}
\end{deluxetable}

\end{document}